# Combining Alchemical Transformation with Physical Pathway to Accurately Compute Absolute Binding Free Energy

8/27/2018

*Nanjie Deng[1]\*, Lauren Wickstrom[2], Emilio Gallicchio[3]*

1. Department of Chemistry and Physical Sciences, Pace University, New York, NY 10038

2. Borough of Manhattan Community College, the City University of New York, Department of Science, New York, NY 10007

3. Department of Chemistry, Brooklyn College, City University of New York, Brooklyn, NY 11210

Email: nanjie.deng@gmail.com

**ABSTRACT**

We present a new method that combines alchemical transformation with physical pathway to accurately and efficiently compute the absolute binding free energy of receptor-ligand complex. Currently, the double decoupling method (DDM) and the potential of mean force approach (PMF) methods are widely used to compute the absolute binding free energy of biomolecules. The DDM relies on alchemically decoupling the ligand from its environments, which can be computationally challenging for large ligands and charged ligands because of the large magnitude of the decoupling free energies involved. On the other hand, the PMF approach uses physical pathway to extract the ligand out of the binding site, thus avoids the alchemical decoupling of the ligand. However, the PMF method has its own drawback because of the reliance on a ligand binding/unbinding pathway free of steric obstruction from the receptor atoms. Therefore, in the presence of deeply buried ligand functional groups the convergence of the PMF calculation can be very slow leading to large errors in the computed binding free energy. Here we develop a new method called AlchemPMF by combining alchemical transformation with physical pathway to overcome the major drawback in the PMF method. We have tested the new approach on the binding of a charged ligand to an allosteric site on HIV-1 Integrase. After 20 ns of simulation per umbrella sampling window, the new method yields absolute binding free energies within ~1 kcal/mol from the experimental result, whereas the standard PMF approach and the DDM calculations result in errors of ~5 kcal/mol and > 2 kcal/mol, respectively. Furthermore, the binding free energy computed using the new method is associated with smaller statistical error compared with those obtained from the existing methods.

# INTRODUCTION

Accurate prediction of binding affinity from the first principles of statistical thermodynamics is not only essential for understanding the physics of molecular recognition [1], but also important for supporting the structure based drug discovery [2]. Among the several statistical thermodynamics based methods developed for computing the absolute binding free energy, including double decoupling method (DDM)[3], the potential of mean force method (PMF),[4] Metadynamics,[5] the Binding energy Distribution method (BEDAM)[6] and Mining Minima (M2),[7] the DDM has been applied to many small to medium sized ligands with reasonably good accuracy. However, because DDM involves alchemically decoupling the ligand from its environments, i.e. from the receptor-ligand complex in solution and from free solution, the method has difficulty in treating the binding of large or charged ligands. For charged ligands, the electrostatic free energies from decoupling the ligand from the complex and that from the solution are both very large and favorable. The precise cancellation of the two very large, opposite terms, which is required to obtain accurate estimate of $\Delta G_{bind}^o$, can be difficult to achieve.[8] Furthermore, because DDM involves charge-charge decoupling between the ligand and its environment, complex corrections may be necessary to account for the errors caused by the use of finite sized, periodic solvent box.[9] In addition to these problems for charged ligands, for large ligands, the convergence of the large Lennard-Jones decoupling free energies can also be very slow and contain large statistical errors.

By using physical pathway to connect the bound state and the unbound state, the PMF approach avoids many of the aforementioned problems in DDM.[8] However, a major problem of the PMF method is that the computation of the PMF depends on the existence of pathways free of major steric clash with the neighboring receptor residues.[8] For larger ligands in more enclosed binding pockets, there is a high probability that certain atoms of the ligand will clash with the receptor atoms around the binding site at some point during the unbinding process. As a result, the convergence of the absolute binding free energy can be very slow in these situations, resulting in very large errors in the apparent absolute binding free energy.[8]

To improve the accuracy and efficiency of binding free energy calculations over the DDM and PMF methods, here we modify the thermodynamic cycle used in the conventional PMF method by introducing alchemical transformations within the physical pathway approach to avoid the potential steric clash between the unbinding ligand the receptor in the course of ligand pulling. After the ligand is pulled out, we add back the free energy cost associated with the alchemical steps in the calculation of the absolute binding free energy. We name the new method AlchemPMF. We have tested the new PMF method on a complex of a HIV-1 Integrase inhibitor carrying negative formal charge and compared the estimated absolute binding free energies from DDM, PMF and the new method. Compared with the existing PMF method and DDM, the new method yields significantly more accurate absolute binding free energy estimates and smaller statistical errors.

**MATERIALS AND METHODS**

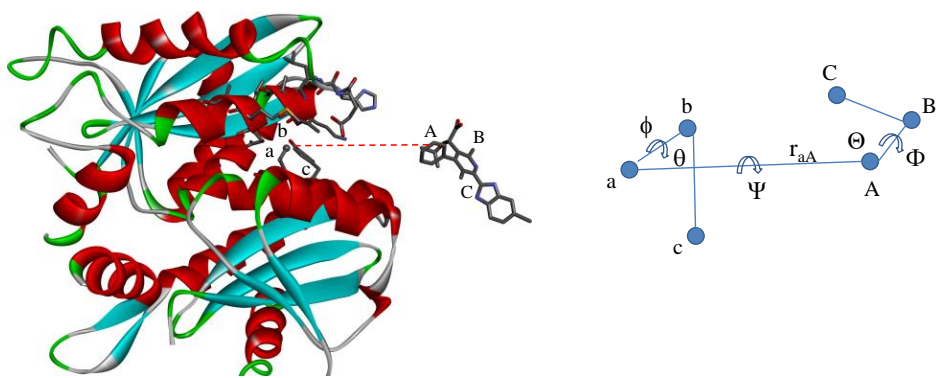

*Figure 1. The coordinate frame in which the orientation and the position of the ligand relative to the receptor are defined. The ligand position is defined in the polar coordinates by the distance $r_{aA}$ and two angles $\theta$: b-a-A; $\phi$: c-b-a-A. The ligand orientation is defined by the three Euler angles: $\Theta$: a-A-B; $\Phi$: a-A-B-C; $\Psi$: b-a-A-B.*

**The simple PMF approach for absolute binding free energy**

The absolute binding free energy $\Delta G_{bind}^{\circ}$ can be written as[3]

$$\Delta G_{bind}^{\circ} = -k_B T \ln \frac{V}{V_0} \frac{Z_{RL,N} Z_{0,N}}{Z_{R,N} Z_{L,N}} = -k_B T \ln \frac{V}{V_0} \frac{Z_{RL,N}}{Z_{R+L,N}} \quad (1)$$

Heree $V_0 = \frac{1}{C^o} = 1660$ Å$^2$ is the inverse of the standard concentration of $C^o = 1$ M solution. $Z_{X,N}$ is the configuration integral ($Z = \int e^{-U(r)/k_B T} dr$) of a single solute X solvated in a box of volume V containing N waters. $Z_{RL,N}$ represents a system containing one receptor-ligand complex in the bound state solvated by N water molecules, $Z_{P+L,N}$ represents a system containing receptor R, an unbound ligand L and N waters, and $Z_{0,N}$ represents a system of N solvent molecules.

In the conventional PMF approach, the right hand side of Eq. (1) is evaluated by inserting intermediate states[8, 10]

$$\Delta G_{bind}^{\circ} = -k_B T \ln \frac{V}{V_0} \frac{Z_{RL,N}}{Z_{R+L,N}} = -k_B T \ln \frac{V}{V_0} \frac{Z_{RL,N}}{Z_{RL(\theta,\phi,\Theta,\Phi,\Psi),N}} \frac{Z_{RL(\theta,\phi,\Theta,\Phi,\Psi),N}}{Z_{R+L(r^*,\theta,\phi,\Theta,\Phi,\Psi),N}} \frac{Z_{R+L(r^*,\theta,\phi,\Theta,\Phi,\Psi),N}}{Z_{R+L,N}} \quad (2)$$

Here $Z_{RL(\theta,\phi,\Theta,\Phi,\Psi),N}$ represents a bound complex RL in which the ligand L's external degrees of freedom, which are defined by the polar angles (θ, φ) and three Euler angles (Θ, Φ, Ψ), are restrained to their equilibrium values in the bound state by a set of harmonic restraints $(U_\theta, U_\phi, U_\Theta, U_\Phi, U_\Psi)$: $U_\theta = \frac{1}{2} k_\theta (\theta - \theta_0)^2$, $U_\phi = \frac{1}{2} k_\phi (\phi - \phi_0)^2$, $U_\Theta = \frac{1}{2} k_\Theta (\Theta - \Theta_0)^2$, $U_\Phi = \frac{1}{2} k_\Phi (\Phi - \Phi_0)^2$, and $U_\Psi = \frac{1}{2} k_\Psi (\Psi - \Psi_0)^2$. Fig. 1 gives the definition of the polar angles (θ, φ) and three Euler angles (Θ, Φ, Ψ). $Z_{R+L(r^*,\theta,\phi,\Theta,\Phi,\Psi),N}$ represents a system in which the ligand L is subject to both the polar and orientational restraints $(U_\theta, U_\phi, U_\Theta, U_\Phi, U_\Psi)$, and to the harmonic restraint $U_{r^*} = \frac{1}{2} k_r (r - r^*)^2$, which forces the ligand atom A to be close to a bulk location r*.

Thus, the different terms inside the logarithm of the right hand side of Eq. (2) correspond to the following thermodynamic transformations (the corresponding thermodynamic cycle is illustrated by Fig. 2):

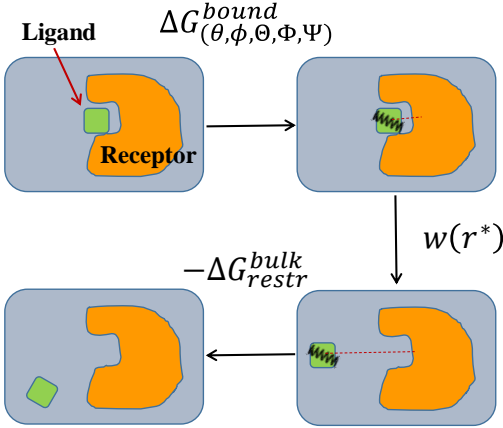

*Figure 2. Thermodynamic pathway used by the standard PMF method for computing absolute binding free energy.*

The first term $\frac{Z_{RL,N}}{Z_{RL(\theta,\phi,\Theta,\Phi,\Psi),N}}$ corresponds to the free energy of switching on the angular restraints $(U_\theta, U_\phi, U_\Theta, U_\Phi, U_\Psi)$ on the ligand when it is in the bound state, i.e.

$$\frac{Z_{RL,N}}{Z_{RL(\theta,\phi,\Theta,\Phi,\Psi),N}} = e^{\Delta G^{bound}_{(\theta,\phi,\Theta,\Phi,\Psi)}/k_B T} \quad (3.1)$$

This term is computed using free energy perturbation (FEP) on the bound complex by computing the free energy change of switching on the restraints $(U_\theta, U_\phi, U_\Theta, U_\Phi, U_\Psi)$.

The second term $\frac{Z_{RL(\theta,\phi,\Theta,\Phi,\Psi),N}}{Z_{R+L(r^*,\theta,\phi,\Theta,\Phi,\Psi),N}}$ equals to the ratio of the probability of an angularly restrained bound complex and the probability of an unbound, but "cross linked" receptor-ligand "complex", in which the cross-linking is maintained by the restraints $(U_{r^*}, U_\theta, U_\phi, U_\Theta, U_\Phi, U_\Psi)$. Let $w(r)$ be the 1D potential of mean force (PMF) along the $r_{aA}$ axis. We have previously shown that[10]

$$\frac{Z_{RL(\theta,\phi,\Theta,\Phi,\Psi),N}}{Z_{R+L(r^*,\theta,\phi,\Theta,\Phi,\Psi),N}} = \frac{\int_{bound} e^{-[w(r)-w(r^*)]/k_B T} dr}{(\frac{2\pi k_B T}{k_r})^{\frac{1}{2}}} \quad (3.2)$$

Here the 1D PMF $w(r)$ is computed in the presence of the angular restraints $(U_\theta, U_\phi, U_\Theta, U_\Phi, U_\Psi)$ from umbrella sampling, in which a series of MD simulations is performed with the harmonic distance restraint between the receptor atom *a* and ligand atom *A*.

The last term $\frac{Z_{R+L(r^*,\theta,\phi,\Theta,\Phi,\Psi),N}}{Z_{R+L,N}}$ represents the free energy of removing all the distance and angular restraints $(U_{r^*}, U_\theta, U_\phi, U_\Theta, U_\Phi, U_\Psi)$ on the bulk ligand such that the ligand is allowed to occupy the standard volume $1/C°$ and rotate freely. When the force constants used in the harmonic restraints are sufficiently strong, the rigid-rotor approximation applies, allowing this term to be evaluated analytically as[10]

$$\frac{Z_{R+L(r^*,\theta,\phi,\Theta,\Phi,\Psi),N}}{Z_{R+L,N}} \approx \frac{r^{*2}\sin\theta_0\sin\Theta_0(2\pi k_B T)^3}{8\pi^2 V(k_r k_\theta k_\phi k_\Theta k_\Phi k_\Psi)^{\frac{1}{2}}} \tag{3.3}$$

where $\theta_0$ and $\Theta_0$ are the equilibrium values of θ and Θ, respectively.

Substituting Eq. (3.1) through (3.3) into Eq. (2), we obtain the expression of $\Delta G_{bind}^\circ$ in the standard PMF approach

$$\Delta G_{bind}^\circ = -\Delta G_{(\theta,\phi,\Theta,\Phi,\Psi)}^{bound} - w(r^*) - k_B T \ln \frac{\int_{bound} e^{-[w(r)-w(r^*)]/k_B T} dr}{(\frac{2\pi k_B T}{k_r})^{\frac{1}{2}}} - k_B T \ln \frac{C^\circ r^{*2}\sin\theta_0\sin\Theta_0(2\pi k_B T)^3}{8\pi^2(k_r k_\theta k_\phi k_\Theta k_\Phi k_\Psi)^{\frac{1}{2}}}$$

(4)

**A new thermodynamic pathway to compute absolute binding free energy: AlchemPMF**

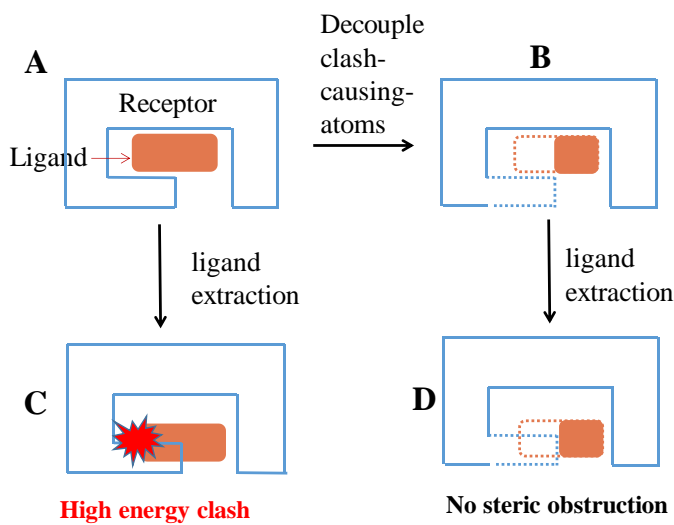

*Figure 3. A→C represents he thermodynamic path in the standard PMF method, while A→B→D. is the thermodynamic path used in the new PMF method.*

In the standard PMF approach, the interactions between the ligand and the receptor are not scaled. In a more enclosed binding pocket, the steric clash between certain atoms in the ligand and the receptor will lead to high energy barriers. Such high energy barriers will artificially inflate the computed value of the PMF in the bulk region $w(r^*)$, since the protein residues involved in the clash will not have sufficient time to spontaneously sample the more open conformation within the tens of nanosecond timescale of the umbrella sampling.

To avoid the interference of the high energy states arising from the atomic clashes, we insert an alchemical transformation in the thermodynamic cycle. We first identify the group of atoms in both the ligand and the receptor that will collide with each other along the unbinding pathway. These atoms are referred as clashing-causing-atoms. Then, in the new thermodynamic cycle (Fig. 3), after switching on the angular restraints $(U_{r^*}, U_\theta, U_\phi, U_\Theta, U_\Phi, U_\Psi)$, we alchemically turn off the pairwise interactions between these two groups of atoms only. Now we follow the standard procedure to pull the ligand out of the binding site along the chosen pathway, which will now be free from obstruction because of the previous alchemical transformation step. In the post-processing step, the free energy of turning off the pairwise interactions between the clash-causing atoms in the ligand and those in the receptor need to be included in calculation of the total absolute binding free energy. Fig. 3 illustrate the alchemical transformations in the new PMF method. Mathematically, this means that the Eq. (2) is modified by inserting two additional intermediate states, i.e.

$$\Delta G^\circ_{bind} = -k_B T \ln \frac{V}{V_0} \frac{Z_{RL,N}}{Z_{R+L,N}} =$$

$$-k_B T \ln \frac{V}{V_0} \frac{Z_{RL,N}}{Z_{RL(\theta,\phi,\Theta,\Phi,\Psi),N}} \frac{Z_{RL(\theta,\phi,\Theta,\Phi,\Psi),N}}{Z_{R'L'(\theta,\phi,\Theta,\Phi,\Psi),N}} \frac{Z_{R'L'(\theta,\phi,\Theta,\Phi,\Psi),N}}{Z_{R'+L'(r^*,\theta,\phi,\Theta,\Phi,\Psi),N}} \frac{Z_{R'+L'(r^*,\theta,\phi,\Theta,\Phi,\Psi),N}}{Z_{R+L(r^*,\theta,\phi,\Theta,\Phi,\Psi),N}} \frac{Z_{R+L(r^*,\theta,\phi,\Theta,\Phi,\Psi),N}}{Z_{R+L,N}}$$

(5)

Here $R'$ and $L'$ represent the alchemically modified receptor and ligand in which the interactions between their clashing-causing-atoms are turned off. Comparing Eq. (5) with Eq. (2), two transformations are added: $\frac{Z_{RL(\theta,\phi,\Theta,\Phi,\Psi),N}}{Z_{R'L'(\theta,\phi,\Theta,\Phi,\Psi),N}}$ corresponds to the free energy change of alchemically switching off the pairwise interactions between clash-causing-atoms in the ligand and those in the receptor while the ligand is in the bound state. $\frac{Z_{R'+L'(r^*,\theta,\phi,\Theta,\Phi,\Psi),N}}{Z_{R+L(r^*,\theta,\phi,\Theta,\Phi,\Psi),N}}$ corresponds to the free energy change of alchemically switching on those interactions turned off in the bound state, when the ligand is unbound. Thus, the resulting expression for the $\Delta G^\circ_{bind}$ becomes

$$\Delta G^\circ_{bind} = -\Delta G^{bound}_{(\theta,\phi,\Theta,\Phi,\Psi)} - \Delta G^{bound}_{decouple} - w(r^*) - k_B T \ln \frac{\int_{bound} e^{-w(r)/k_B T} dr}{(\frac{2\pi k_B T}{k_r})^{\frac{1}{2}}} - \Delta G^{bulk}_{couple} -$$

$$k_B T \ln \frac{C^\circ r^{*2} \sin\theta_0 \sin\Theta_0 (2\pi k_B T)^3}{8\pi^2 (k_r k_\theta k_\phi k_\Theta k_\Phi k_\Psi)^{\frac{1}{2}}} \tag{6}$$

where $\Delta G^{bound}_{decouple} = -k_B T \ln \frac{Z_{RL(\theta,\phi,\Theta,\Phi,\Psi),N}}{Z_{R'L'(\theta,\phi,\Theta,\Phi,\Psi),N}}$ is the free energy of alchemically decoupling those clashing-causing-atoms in the bound state, and $\Delta G^{bulk}_{couple} = -k_B T \ln \frac{Z_{R'+L'(r^*,\theta,\phi,\Theta,\Phi,\Psi),N}}{Z_{R+L(r^*,\theta,\phi,\Theta,\Phi,\Psi),N}}$ is the free energy of recoupling those clashing-causing-atoms when the ligand is unbound. Both terms are readily computed using FEP or TI.

**PMF calculation setup**

We compute the ligand 1D PMF $w(r)$ using umbrella sampling simulations in explicit solvent, in which a series of MD simulations is performed with the harmonic distance restraint on the receptor atom $a$ and ligand atom $A$: see Fig. 1. The biasing potential in the $i$-th simulation window is $U_{r,i} = \frac{1}{2}k_r(r - r_i^0)^2$, where $r_i^0$ is the reference distance for the $i$-th sampling window. The full range of the distance space is covered using between 20 and 24 umbrella windows for the different complexes. A single force constant $k_r$ = 1000 kJ mol$^{-1}$ nm$^{-2}$ is used for the distance restraint in all the sampling windows. The force constants used in the angular restraints are: $k_\theta = k_\phi = k_\Theta = k_\Phi = k_\Psi = 1000\ kJ\ rad^{-1}mol^{-1}$. In each umbrella window, a 30-ns MD simulation is performed, starting from the last simulation snapshot of the previous window. The biased probability distributions along the distance $r$ accumulated in these sampling windows are unbiased and combined using the Weighted Histogram Analysis Method (WHAM) method to yield the unbiased distribution and the potential of mean force $w(r)$ The WHAM program implemented by Grossfield is used to calculate the PMF. Five independent umbrella sampling simulations were preformed and the statistical uncertainties in the calculated PMF are estimated by the standard deviation of the results obtained from these independent umbrella sampling simulations. The total simulation time used to compute the PMF for a single receptor-ligand complex is ~2.1 μs.

The paths along which ligands are pulled out in the PMF calculation are identified by trial and error. First, a ligand atom **A** and a protein atom **a** are chosen to define axis $r_{aA}$; using the Discovery Studio Visualizer (Biovia Inc), the ligand is translated manually along this axis with its orientation relative to the receptor fixed. If the unbinding ligand collides with the nearby protein atoms, (a clash is shown as red dashed line in the DS Visualizer), then a different set of atom pairs will be tried. After a few tries, one or more low energy paths may be identified; these paths are then used to run umbrella-sampling simulation to get PMF. For BI-224436, we identified two low free energy paths, with the $\Delta G^{\circ}_{bind}$ estimated using path 2 lower than that from path 1. The PMF result reported here are for the lowest free energy paths.

**MD Simulation Setup**

In this work, the MD free energy simulations were performed using the GROMACS 4.6.4.[11] The atomic coordinates of the bound complex of HIV-1 integrase and the ligand BI-224436 was obtained using the Glide docking program from the Schrodinger Inc..[12] BI-224436 is known to bind at the catalytic core dimer interface of HIV-1 integrase.[13] Therefore, the crystal structure of the HIV-integrase in complex with a similar ligand BI-D[14] is used as the template for Glide docking of the BI-224436. The AMBER parm99ILDN force field[15] is used to model the HIV-1 integrase catalytic core dimer, and the Amber GAFF parameters set[16] and the AM1-BCC charge model[17] are used to describe the ligand BI-224436. TIP3P water[18] boxes previously equilibrated at 300 K and 1 atm pressure were used to solvate the protein-ligand complex. The dimension of the solvent box is set up to ensure that the distance between solute atoms from nearest walls of the box is at least 10 Å. To maintain charge neutrality, 1 Cl$^-$ and 1 Na$^+$ are added to the solvent boxes containing the protein-ligand complex and that containing the ligand, respectively. The electrostatic interactions were computed using the particle-mesh Ewald (PME)[19] method with a real space cutoff of 10 Å and a grid spacing of 1.0 Å. MD simulations were performed in the NPT ensemble with a time step of 2 fs.

# RESULTS AND DISCUSSIONS

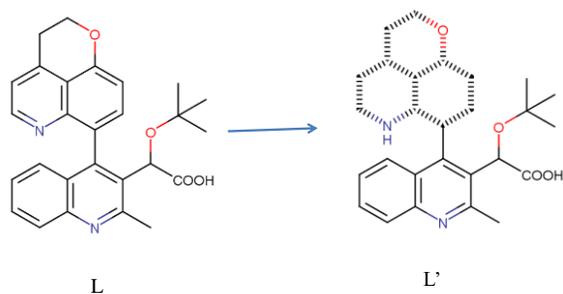

L     L'

*Figure 4.The allosteric HIV-1 Integrase inhibitor BI-224436. Left: the original ligand. Right: a mutated form in which the bulky tricyclic group is decoupled from the environment.*

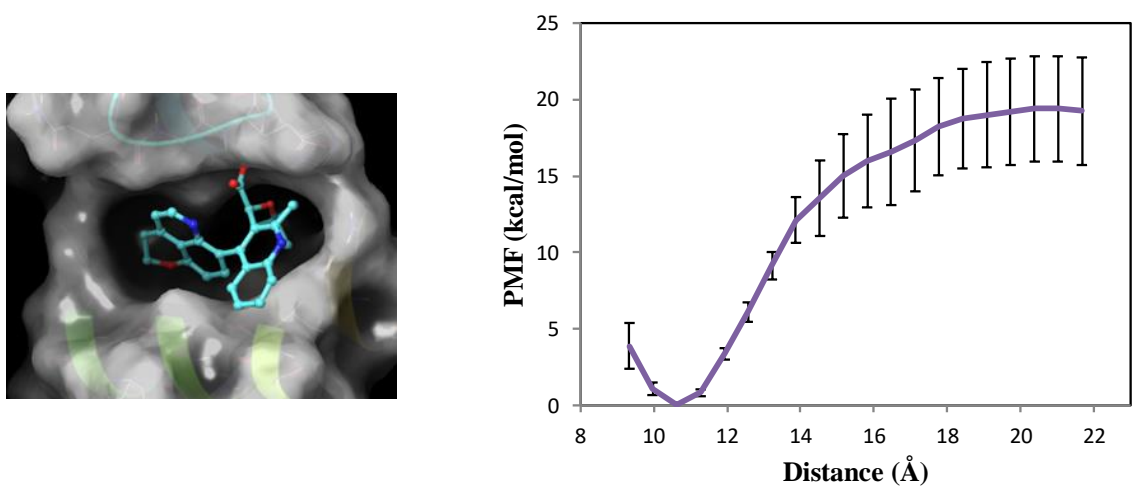

*Figure 5. Left: The charged ligand BI-224436 in the allosteric site of the HIV-1 Integrase catalytic core domain dimer. Right: The PMF of pulling the unmodified BI-224436 out of the binding pocket of HIV-1 integrase.*

We test the new AlchemPMF method by applying it to compute the absolute binding free energy of an allosteric HIV-1 Integrase inhibitor BI-224436 (Fig.2), which carries a negative charge at pH 7. In this ligand, the calculation of $\Delta G°_{bind}$ by using the regular PMF method Eq. (4) is hampered by the presence of the bulky tricyclic group, which clashes with the side chain residues around the binding pocket when the ligand unbinds (Fig.4 and Fig. 5). The steric clash experienced by the unbinding ligand led to large error bars in the calculated PMF of pulling the unmodified BI-224436 out of the binding pocket of HIV-1 integrase: see Fig. 5, right panel. As a result, the $\Delta G°_{bind}$ computed using the simple PMF converges slowly: even with 30 ns of simulation in each of the umbrella sampling window (total simulation time: 24

× 30 = 720 ns) the PMF-calculated $\Delta G°_{bind}$ is still about -3.3 kcal/mol more favorable than the experimental result from SPR measurement: see Table 1 and Fig. 6.

Using the new method AlchemPMF, we first alchemically modify the ligand in the bound state by turning off the nonbonded interaction between the tricyclic group of the ligand and its environment, and then using umbrella sampling to pull the modified ligand out of the binding pocket. Note that other interactions are not modified, including the ligand intramolecular interaction with the tricyclic group. When the ligand is in the bulk solution, we re-couple the disappeared tricyclic group to its environment. Finally, the $\Delta G°_{bind}$ is computed by using the Eq. (6). As seen from Table 1 and Fig. 6, the $\Delta G°_{bind}$ computed from the new procedure converges significantly faster to the experimental result. At t = 20 ns for each of the umbrella sampling windows, the $\Delta G°_{bind}$ computed using Eq. (6) is within 1.2 kcal/mol from the experiment, whereas that the result from regular PMF deviates from the experimental result by ~ 5.2 kcal/mol. Fig. 7 shows the PMF computed using the alchemically modified ligand L', i.e. the right panel of Fig. 4. Compared with the computed PMF for the unmodified ligand (Fig. 5, right panel), the error bars in the PMF function for the alchemically modified ligand is substantially smaller: see Fig. 7. Clearly, by alchemically "hiding" the clash-causing-atoms in the ligand, the steric clash with the receptor in the course of ligand pulling has been largely avoided, which leads to significantly improved binding free energy estimates and smaller statistical errors.

Table 1. Comparisons of the absolute binding free energy for BI-224436 from different methods and using different simulation time segments. Unit: kcal/mol.

| $\Delta G^o_{bind}$ at different time segments[a] | simple PMF[a] | AlchemPMF[b] | DDM | Experiment[c] |
|---|---|---|---|---|
| 0 -10 ns | -18.14 ± 0.89 | -15.51 ± 1.76 | | |
| 10 - 20 ns | -15.35 ± 1.25 | -11.38 ± 1.17 | -12.54 ± 0.3 | -10.33 |
| 20 - 30 ns | -13.75 ± 2.48 | -10.33 ± 1.06 | | |

a. The $\Delta G^o_{bind}$ obtained using different simulation time segments in each umbrella sampling window for simple PMF and AlchemPMF.

b. Computed using Eq. (4) in the Method.

c. Computed using Eq. (6) in the Method.

d. From SPR measurement, to be published result from Mamuka Kvaratskhelia lab.

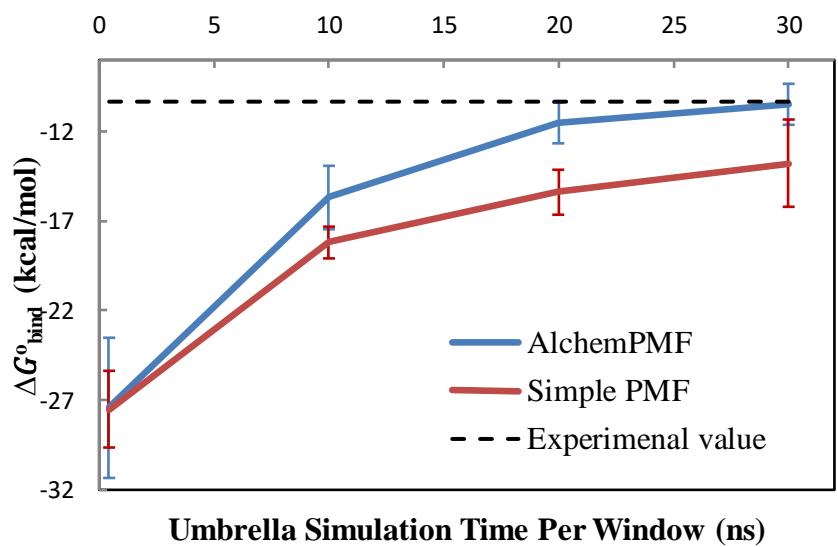

Figure 6. Computed absolute binding free energy as a function of the umbrella sampling simulation time using the simple PMF approach Eq. (4) and AlchemPMF expression Eq. (6).

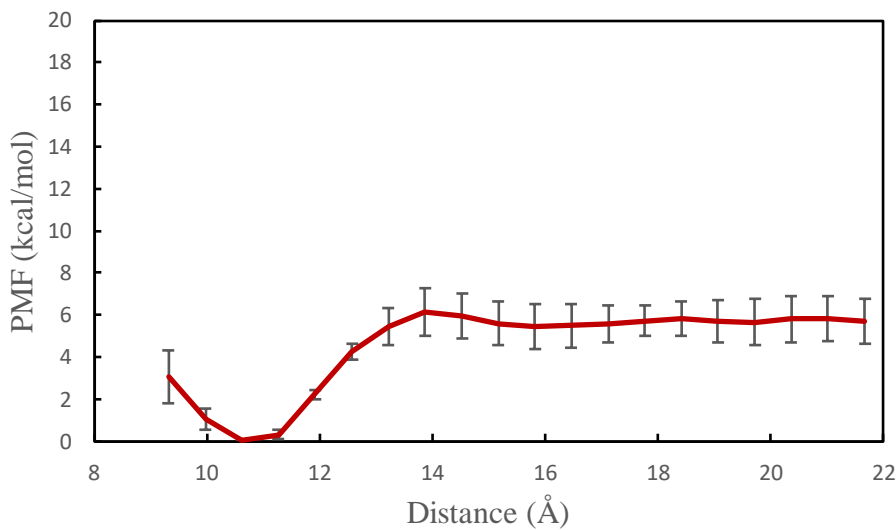

*Figure 7. The PMF of pulling the alchemically modified BI-224436 out of the binding pocket of HIV-1 integrase.*

**CONCLUSIONS**

The PMF approach is a powerful method for computing absolute binding free energy [4, 20], especially in treating binding problems involving charged ligands [8, 10]. In practice, however, the PMF method is not suitable when the binding site is significantly enclosed; this is because a converged calculation of the PMF depends on the existence of a ligand extraction pathway that is free of steric obstruction from receptor atoms. To overcome this major limitation in the standard PMF method, we develop a novel method called AlchemPMF, by combining alchemical transformation with physical pathway to remove steric obstruction in the calculation of PMF. Our test results on the binding of a charged HIV-1 integrase ligand with a buried functional group demonstrates that the AlchemPMF leads to significantly improved absolute binding free energy estimates and smaller statistical errors compared with those given by the simple PMF method and DDM. The new method is expected to allow a broader range of binding problems to be treated accurately, including protein-protein and protein-DNA complexes, which involve larger and more complex binding interfaces compared with those in the protein-small molecule complexes.